\documentclass[conference]{IEEEtran}
\IEEEoverridecommandlockouts
\usepackage{pifont}
\usepackage{amsmath}
\usepackage{amssymb}
\usepackage{amsfonts}
\usepackage{graphicx}
\usepackage{textcomp}
\usepackage{xcolor}
\usepackage{multirow}
\usepackage{soul}
\usepackage{tikz}
\usepackage{booktabs}
\usepackage{algorithm}
\usepackage{algpseudocode}
\usepackage{colortbl}
\usepackage{bm}
\usepackage{float}
\usepackage{filecontents}
\usepackage{CJK}
\usepackage{makecell}
\usepackage{algorithm}
\usepackage{algorithmicx}
\usepackage[a4paper, total={184mm,239mm}]{geometry}
\def\BibTeX{{\rm B\kern-.05em{\sc i\kern-.025em b}\kern-.08em
    T\kern-.1667em\lower.7ex\hbox{E}\kern-.125emX}}
\begin{document}

\title{Compromising the Intelligence of Modern DNNs: On the Effectiveness of Targeted RowPress \vspace{-0.5em}}
\author{Ranyang Zhou$^\dagger$, Jacqueline T. Liu$^\ddagger$, Sabbir Ahmed$^\ddagger$, Shaahin Angizi$^\dagger$, and Adnan Siraj Rakin$^\ddagger$  \\
\small $^\dagger$Department of Electrical and Computer Engineering, New Jersey Institute of Technology, Newark, NJ, USA\\
$^\ddagger$Department of Computer Science, State University of New York at Binghamton, NY, USA\\
shaahin.angizi@njit.edu, arakin@binghamton.edu \vspace{-2em}
\\}
\maketitle

\begin{abstract}
Recent advancements in side-channel attacks have revealed the vulnerability of modern Deep Neural Networks (DNNs) to malicious adversarial weight attacks. The well-studied RowHammer attack has effectively compromised DNN performance by inducing precise and deterministic bit-flips in the main memory (e.g., DRAM). Similarly, RowPress has emerged as another effective strategy for flipping targeted bits in DRAM. However, the impact of RowPress on deep learning applications has yet to be explored in the existing literature, leaving a fundamental research question unanswered: How does RowPress compare to RowHammer in leveraging bit-flip attacks to compromise DNN performance? This paper is the first to address this question and evaluate the impact of RowPress on DNN applications. We conduct a comparative analysis utilizing a novel DRAM-profile-aware attack designed to capture the distinct bit-flip patterns caused by RowHammer and RowPress. Eleven widely-used DNN architectures trained on different benchmark datasets deployed on a Samsung DRAM chip conclusively demonstrate that they suffer from a drastically more rapid performance degradation under the RowPress attack compared to RowHammer. The difference in the underlying attack mechanism of RowHammer and RowPress also renders existing RowHammer mitigation mechanisms ineffective under RowPress. As a result, RowPress introduces a new vulnerability paradigm for DNN compute platforms and unveils the urgent need for corresponding protective measures.
\end{abstract}

\section{Introduction}
Recent advancements in deep learning have revolutionized applications, including image classification \cite{img_clf}, object detection \cite{dfr_tsd}, and speech recognition \cite{xiong2016achieving}. However, privacy and security concerns surrounding this powerful technology are gaining increasing attention, especially in safety-critical domains like healthcare and finance \cite{rajabli2020software}. Recent attacks exploiting software \cite{gu2019badnets,madry2018towards} and system-level \cite{yao2020deephammer,hong2019terminal} vectors demonstrate the feasibility of compromising DNN systems.

Among the various security threats to DNN security, this paper focuses on \textit{adversarial weight attack} \cite{rakin2019bit,yao2020deephammer,rakin2020tbt,hong2019terminal,chen2021proflip}. This attack typically injects RowHammer-based faults \cite{kim2014flipping} in DRAM, where model parameters are stored. RowHammer, characterized by repetitive DRAM row activation (i.e., hammering), can significantly degrade DNN performance through precise bit-flip algorithms \cite{rakin2019bit}. Advanced in-DRAM defenses \cite{seyedzadeh2018mitigating,saroiu2022price,kim2014architectural,qureshi2022hydra} have improved memory system resilience, mitigating RowHammer threats, but similar emerging threats, like RowPress, remain a concern.

RowPress shares similarities with RowHammer in exploiting the differential bit states between adjacent rows. The fundamental difference between them lies in the \textit{underlying attack mechanism}. RowPress relies on prolonged row activation, as opposed to RowHammer's frequent and relatively transient row activation, and thus requires considerably fewer number of activations to induce bit-flips. \cite{luo2023rowpress} also shows that these two attack models result in distinct bit-flip patterns, i.e., vulnerable bit profiles. Existing precise bit-flip algorithms such as \cite{rakin2019bit} were designed according to RowHammer. Given the aforementioned similarities and differences between RowHammer and RowPress, we are motivated to leverage and revise such algorithm by incorporating two distinct vulnerable bit profiles to compare and analyze the effect of these two attack models on depleting DNN intelligence. The main contributions of this work are as follows:

\begin{enumerate}
    \item We perform RowHammer \cite{mutlu2023fundamentally} and RowPress \cite{luo2023rowpress} attacks on a Samsung-manufactured DDR4 DRAM chip to profile vulnerable bit-cell locations under both attacks. We then develop a novel profile-aware bit-flip attack algorithm that utilizes these two profiles to induce precise targeted bit-flips.
    \item We conduct an extensive comparative analysis between RowHammer and RowPress on their effectiveness in depleting DNN intelligence. Eleven DNNs of different sizes and structures trained on three benchmark datasets, including both image and speech data modality, are experimented.
    \item Our results exhibit that, compared to RowHammer, RowPress only requires up to \emph{4 $\times$} fewer bit-flips to undermine DNN performance and can induce up to \emph{20 $\times$} more bit-flips within the same operational duration, making it a stealthier attack with noticeably higher efficacy. 
\end{enumerate}

\section{Background}
\noindent\textbf{DRAM Organization \& Commands.} A DRAM chip consists of multiple memory banks, each comprising 2D sub-arrays of memory cells arranged in matrices, with billions of cells on modern chips. Each DRAM cell contains a capacitor and an access transistor, where the capacitor's charge state represents a binary '1' or '0' \cite{zhou2022red,angizi2019graphide}. In idle mode, the memory controller issues a Precharge (\texttt{PRE}) command, precharging the Bit-Line (BL) to $\frac{V_{DD}}{2}$. In active mode, the Activate (\texttt{ACT}) command activates the Word-Line (WL), allowing DRAM cells to share charge with the BL, altering its voltage. A sense amplifier detects this deviation and amplifies it to $V_{DD}$ or 0, enabling data transfer via read (\texttt{RD}) / write (\texttt{WR}) commands \cite{zhang2023aligner,zhou2022flexidram}.

\noindent\textbf{DRAM Timing Parameters.} The most basic parameter is the clock cycle ($t_{CK}$) that is used to measure all parameters. Row Active Time ($t_{RAS}$) encompasses the temporal window between an \texttt{ACT} command and the subsequent \texttt{PRE} command, ensuring optimal performance by restoring charge within the DRAM cells on the open row. Row Precharge Time ($t_{RP}$) signifies the gap between a \texttt{PRE} command and the next \texttt{ACT} command, closing the open WL and initiating the pre-charging of the BLs to $\frac{V_{DD}}{2}$. Retention time refers to the duration a memory cell can hold its stored data without needing a refresh, influenced by factors like cell density and electromagnetic interference. These parameters ensure reliable and efficient DRAM operation. In the RowHammer model, the retention time of certain victim rows may significantly reduce. The Refresh Window ($t_{REFW}$) is the interval within which all DRAM cells must be refreshed to prevent data loss or corruption \cite{angizi2019redram}.

\begin{figure}[t]
\begin{center}
\begin{tabular}{c}
\includegraphics [width=0.94\linewidth]{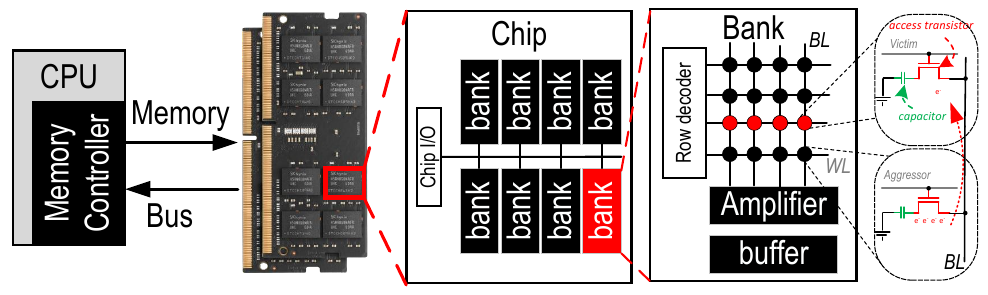}\vspace{-0.4em}
 \end{tabular}
\caption{DRAM organization with fault injection model}\vspace{-2.1em}
\label{DRAM}
\end{center}
\end{figure}

\noindent\textbf{RowHammer in DDR4 \& Protection Mechanisms.} Kim et al. \cite{kim2014flipping} conducted an extensive study on RowHammer bit-flips in DDR3 modules, finding that approximately 85\% of the tested modules were susceptible to RowHammer attacks. Thus, earlier RowHammer research focused on DDR3 systems \cite{seaborn2015exploiting}. DDR4 modules, introduced to create a RowHammer-less landscape, have documented instances of RowHammer on earlier DDR4 generations \cite{lipp2020nethammer,gruss2018another}. TRRespass \cite{frigo2020trrespass} is the only recently established work exploring the multi-sided fault injection model. Multiple software and hardware mitigation mechanisms reduce RowHammer-based attacks \cite{kim2014flipping,marazzi2022protrr,zhou2023dnn,zhou2024assessing}. Hardware-based research is categorized into \textit{victim-focused} mechanisms with probabilistic refreshing (e.g., PRA \cite{kim2014architectural}, ProTRR \cite{marazzi2022protrr}) or non-probabilistic methods \cite{zhou2024dnn,ranyang2023ppim,zhou2024dram} and \textit{aggressor-focused} mechanisms counting activations (e.g., TRR \cite{hassan2021uncovering}, Hydra \cite{qureshi2022hydra}, CBT \cite{seyedzadeh2018mitigating}, Panopticon \cite{bennett2021panopticon}, CRA \cite{kim2014architectural}, TWiCe \cite{lee2019twice}, Graphene \cite{park2020graphene}, Mithril \cite{kim2022mithril}). System manufacturers tend to follow mechanisms that detect RowHammer conditions and intervene, such as increasing refresh rates and counter-based approaches. TRR \cite{frigo2020trrespass} and counter-based detection methods \cite{kim2014architectural,qureshi2022hydra,seyedzadeh2016counter} require additional hardware to calculate and record row activations to other fast-read-memory (SRAM \cite{lee2019twice} / CAM \cite{park2020graphene}). The controller refreshes the target row if the number reaches the Maximum Activation Count (MAC) \cite{frigo2020trrespass}. Counter-based solutions add a new DRAM command called Nearby Row Refresh (NRR) \cite{lee2019twice,park2020graphene}, issued to refresh the relevant victim rows. JEDEC standards outline three MAC configurations: (1) unlimited, if the DRAM module is RowHammer-free; (2) untested, if the module has not undergone post-production inspection; or (3) $T_{MAC}$, the specific number of \texttt{ACT}s the module can withstand (e.g., 1M). Most DDR4 modules assert an unlimited MAC value \cite{frigo2020trrespass}.

\noindent\textbf{RowPress.} RowPress \cite{luo2023rowpress} and RowHammer share the potential for bit-flips in victim rows when bits differ from adjacent (aggressor) rows. Both attacks get worse as DRAM technology scales down to smaller node sizes. The distinction between them lies in how the DRAM rows are accessed (i.e., the failure mechanism): RowHammer repeatedly opens (i.e., activates) and closes an aggressor row many times, whereas RowPress extends the activation duration in the aggressor row, reducing the required number of activations. RowPress-vulnerable cells and RowHammer-vulnerable cells bear $< .5\%$ overlap and they exhibit opposite bit-flip directionality trends. 

\section{Motivation}
Existing RowHammer protection mechanisms are designed to monitor the frequency of memory row activations within a specified period. Given that RowPress induces bit-flips in a different manner -- prolonging a single activation duration, it is obvious that those RowHammer protection mechanisms will have no effect against RowPress. Although developing a new defense mechanism against RowPress is certainly a meaningful research topic, it is not the focus of this paper; rather, we would like to first conduct a novel systematic comparative analysis on these two fault injection models, focusing on their impact on DNN applications. To be more specific, we focus on answering the following two research questions:

\vspace{0.5em}
\hspace{-0.5em}\fcolorbox{black}{white}{\begin{minipage}{24em}
\textbf{Research Question 1.} \emph{How effective is RowPress in depleting the intelligence of deep learning models?} 
\end{minipage}}

\vspace{0.5em}
\hspace{-0.5em}\fcolorbox{black}{white}{\begin{minipage}{24em}
\textbf{Research Question 2.} \emph{Is it RowHammer or Rowpress that is more effective in performing targeted bit-flip attacks to compromise DNN performance?}
\end{minipage}}\vspace{0.5em}

We hope that our results will motivate the research community in brainstorming effective defense mechanisms against RowPress.

\section{Threat Model}
We adopt a standard practical threat model similar to prior works leveraging remote side-channel attacks to compromise DNN performance \cite{yao2020deephammer,rakin2022deepsteal,hong2019terminal,olgun2021quac,rakin2019bit,chen2021proflip,rakin2021deep,bai2023versatile}. We assume that DNN model inference takes place in a resource-sharing environment as in the Machine-Learning-as-a-Service (MLaaS) \cite{ribeiro2015mlaas} setting. The attacker can run user-level un-privileged processes remotely on the same machine where the victim DNN model is deployed. The attacker can map the virtual addresses to physical addresses using several techniques such as leveraging huge page support, hardware-based side-channel attack \cite{gruss2018another}, and memory messaging \cite{kwong2020rambleed}. The attacker requires knowledge of the DRAM memory addressing scheme, which can be obtained via reverse-engineering \cite{pessl2016drama}. We assume the attacker can engender targeted bit-flips at desired locations using fast and precise multi-bit-flip techniques \cite{yao2020deephammer} that ensure the correct hammering patterns for RowHammer or keep a targeted row open for RowPress. Nevertheless, we assume the kernel and operating system are trusted and well-protected \cite{konoth2018zebram}. Also following standard practices, we assume ECC does not protect the commercial DRAM and cannot protect large-scale deep learning models against RowHammer \cite{yao2020deephammer,rakin2022deepsteal}, i.e., the attacker can bypass the tracking algorithm and prevent the triggering of refresh operations. Finally, we assume a white-box attacker for deep learning models, following earlier works on adversarial weight attacks \cite{rakin2019bit,hong2019terminal,yao2020deephammer,chen2021proflip,rakin2021deep}. Recent remote side-channel attacks have made this white-box threat model assumption practical, as attackers can now effectively obtain critical information such as the number of layers, layer sizes, weight bit sizes, and parameter values through remote side-channel methods \cite{244042,xiang2020open,yu2020deepem,rakin2022deepsteal}. However, the attacker does not have access to training information such as datasets, hyperparameters, etc.

\section{Fault Injection Models}
In this section, we detail the two fault injection models: RowHammer and RowPress, as well as how we obtain the vulnerable bit-flip profiles.

\begin{figure}[t]
\begin{center}
\begin{tabular}{c}
\includegraphics [width=0.71\linewidth]{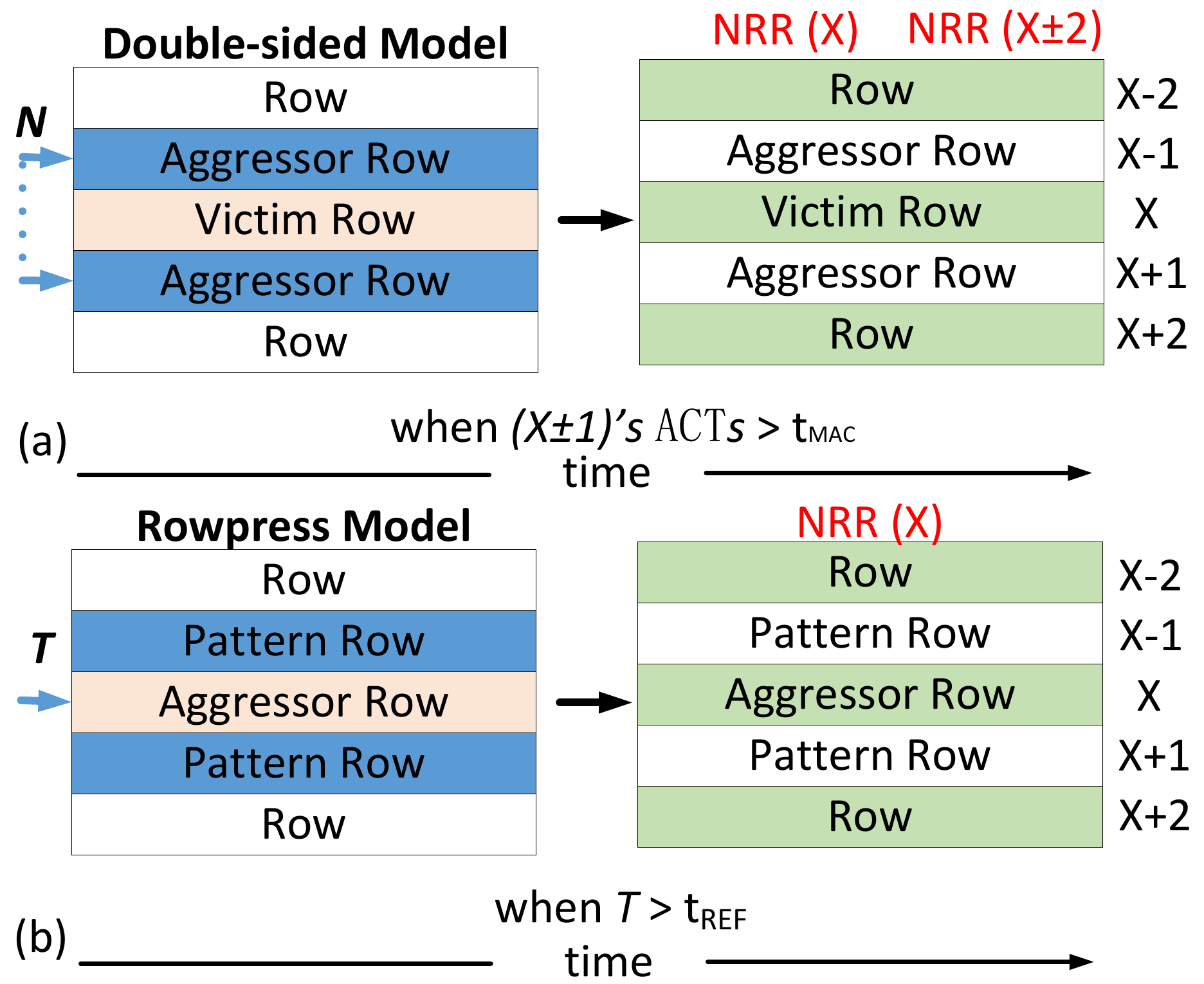}
 \end{tabular} 
\caption{(a) Double-sided RowHammer attack model, (b) RowPress attack model.}
\label{RHmodel}
\end{center}\vspace{-1.5em}
\end{figure}
\subsection{RowHammer}
Traditional fault injection models such as double-sided
attacks can be effectively defended by counter-based frameworks \cite{saroiu2022price,kim2014architectural,qureshi2022hydra,seyedzadeh2016counter}. As shown in Fig. \ref{RHmodel} (a), the double-sided RowHammer model mainly affects the victim rows with two aggressor rows X$\pm$1. We manually insert data patterns into three rows, assigning all 1s to two aggressors X$\pm$1 and all 0s to the victim row X. This is intended to simulate the ideal case where all bits in the victim row differ from those in the aggressor rows. Subsequently,  \texttt{ACT} command is continuously issued to the aggressor rows. 
The following testing allows us to establish a range of aggressor rows' HCs, denoted by $N$, which effectively quantifies the vulnerability level of the victim row. The lower and higher bounds of $N$ correspond to the respective thresholds where the victim row first exhibits bit-flips and where the victim row is entirely reversed due to the attacks. Hence, defense mechanisms will easily identify aggressor rows that have been activated significantly more frequently than other normal rows. As discussed, such defenses establish distinct thresholds depending on the manufacturer of the chips. If the defense mechanism properly detects that the row X$\pm$1 reaches the $MAC$, the NRR will refresh row X and X$\pm$2 as shown in Fig. \ref{RHmodel}(a). 
Fig. \ref{timing} (a) shows the timing for such a RowHammer attack. Assuming RowHammer is implemented on row 0x99. F is a flag used to decide whether to issue an NRR command or not. When HC of the row surpasses MAC, which means $t_{RAS}$$\times$ HC $\geq$$T_{MAC}$, the memory controller will consider an NRR operation for that row. Algorithm \ref{alg-RH} provides a visible function to understand the flow. Firstly, we define data$\_$pattern as all `1's and data$\_$pattern$\_$inv as all `0's in lines 5 and 6, respectively, to create the ideal conditions for a RowHammer attack. Subsequently, in lines 7 and 8, we write these patterns into the aggressor rows and the victim row. Then, the loop operation described from lines 9 to 12 simulates the attacker continuously activating the aggressor rows. Finally, in lines 16 to 18, we transfer the data from the chips to the host PC and check for bit-flips.

\begin{algorithm}[t]
        \caption{\small RowHammer Fault Injection}
          \scalebox{0.65}{
    \begin{minipage}{3\linewidth}
        \begin{algorithmic}[1]
            \State $\textbf{Procedure: \textit{RowHammer}}$
                \State $\textbf{Input} \hspace{4pt} N$ 
                \State $Allocate \hspace{4pt} row\_address, bank\_address, column\_address$ 
                \State $Load \hspace{4pt} Data\_pattern \hspace{4pt} $\&$ \hspace{4pt} Data\_pattern\_inv $ 
                \State $Aggressor\_row[row] \gets Data\_pattern\hspace{16pt}$
                \State $Victim\_row[row] \gets Data\_pattern\_inv\hspace{16pt}//We\hspace{4pt}assign\hspace{4pt}0x00000000\hspace{4pt}to\hspace{4pt}victim\hspace{4pt}rows$ 
                \State $\textbf{For}\hspace{4pt}(RowHammer\_cnt < N)\hspace{4pt}\textbf{do}$
                    \State $\hspace{16pt}\textbf{For}\hspace{4pt}(row\hspace{4pt} in\hspace{4pt}Row\_address)\hspace{4pt}\textbf{do}$
                    \State $\hspace{32pt}ACT \hspace{4pt} Aggressor\_row[row];\hspace{16pt}//Keep\hspace{4pt}hammering\hspace{4pt}Row\hspace{4pt}X\pm1$ 
                    \State $\hspace{32pt}PRE \hspace{4pt} Aggressor\_row[row];$ 
                    \State $\hspace{32pt}row \gets Row\_address+1;$ 
                \State $\hspace{16pt}RowHammer\_cnt+1;$
        \State $\textbf{For}\hspace{4pt}(row\hspace{4pt} in\hspace{4pt}Row\_address)\hspace{4pt}\textbf{do}$
                    \State $\hspace{16pt}READ \hspace{4pt} Rows[row];$ 
                    \State $\hspace{16pt}PRE \hspace{4pt} Rows[row];$ 
                    \State $\hspace{16pt}row \gets row+1;$ 
        \State $Receive\_Data(Platform);\hspace{16pt}//Write\hspace{4pt}data\hspace{4pt}back\hspace{4pt}to\hspace{4pt}hostPC$
        \State $Detect\_BitFlips(Victim\_Row)$

    \State \textbf{end} $\textbf{Procedure}$
        \end{algorithmic} 
         \end{minipage} \vspace{-3em}}
         \label{alg-RH}
    \end{algorithm}

Common $t_{RAS}$ values for DDR4 memory modules range from around 36 to 48 $t_{CK}$ \cite{choi2020reducing}, but these values can differ based on the module's speed rating (e.g., DDR4-2133, DDR4-2400, DDR4-3200, etc.). The duration of a clock cycle for DDR4-2400 memory can be calculated as $t_{CK}=\frac{1}{2400M/s}$. In our design, every $t_{RAS}$ consists of three parts: \texttt{ACT}, \texttt{Sleep(S)}, and \texttt{PRE}, where \texttt{Sleep(S)} is set to 5{$t_{CK}$}. Previous research \cite{lang2023blaster} has indicated that the maximum number of HCs typically reaches approximately 1.36 million.

\subsection{RowPress}
RowPress \cite{luo2023rowpress} aims to bypass the counter-based defense such as CAT \cite{seyedzadeh2018mitigating}. Although most counter-based mechanisms require additional hardware resources to mitigate RowHammer attacks, the threshold for triggering alerts is maintained at an extremely low level. This makes it challenging for attackers to compromise the chips and even generate bit-flips. RowPress extends the open window during activation to prolong charge leakage, thereby generating bit-flips, rather than repeatedly activating rows.
Our RowPress implementation is slightly different from \cite{luo2023rowpress}. The only difference lies in the attacking target: instead of keeping the aggressor rows open in Fig. \ref{RHmodel}(a) for a long period, we do so in the victim row in Fig. \ref{RHmodel}(a). This effectively transforms the victim rows into aggressor rows by directly attacking them. Within this context, we propose to designate the rows adjacent to the aggressor rows as pattern rows and the victim row as the aggressor row, as illustrated in Fig. \ref{RHmodel}(b). A key similarity shared between RowHammer and RowPress is that bit-flips occur only when the bits in a victim row differ from those in its adjacent rows. The pattern rows serve as victim rows but are utilized to monitor bit-flip occurrences. Here, $T$, whose value we can customize, represents the open window duration of the victim row. Please note that $T$ cannot exceed the limitation imposed by the refresh time $t_{REF}$.

\begin{figure}[t]
\begin{center}\vspace{-1.4em}
\begin{tabular}{c}
\includegraphics [width=0.99\linewidth]{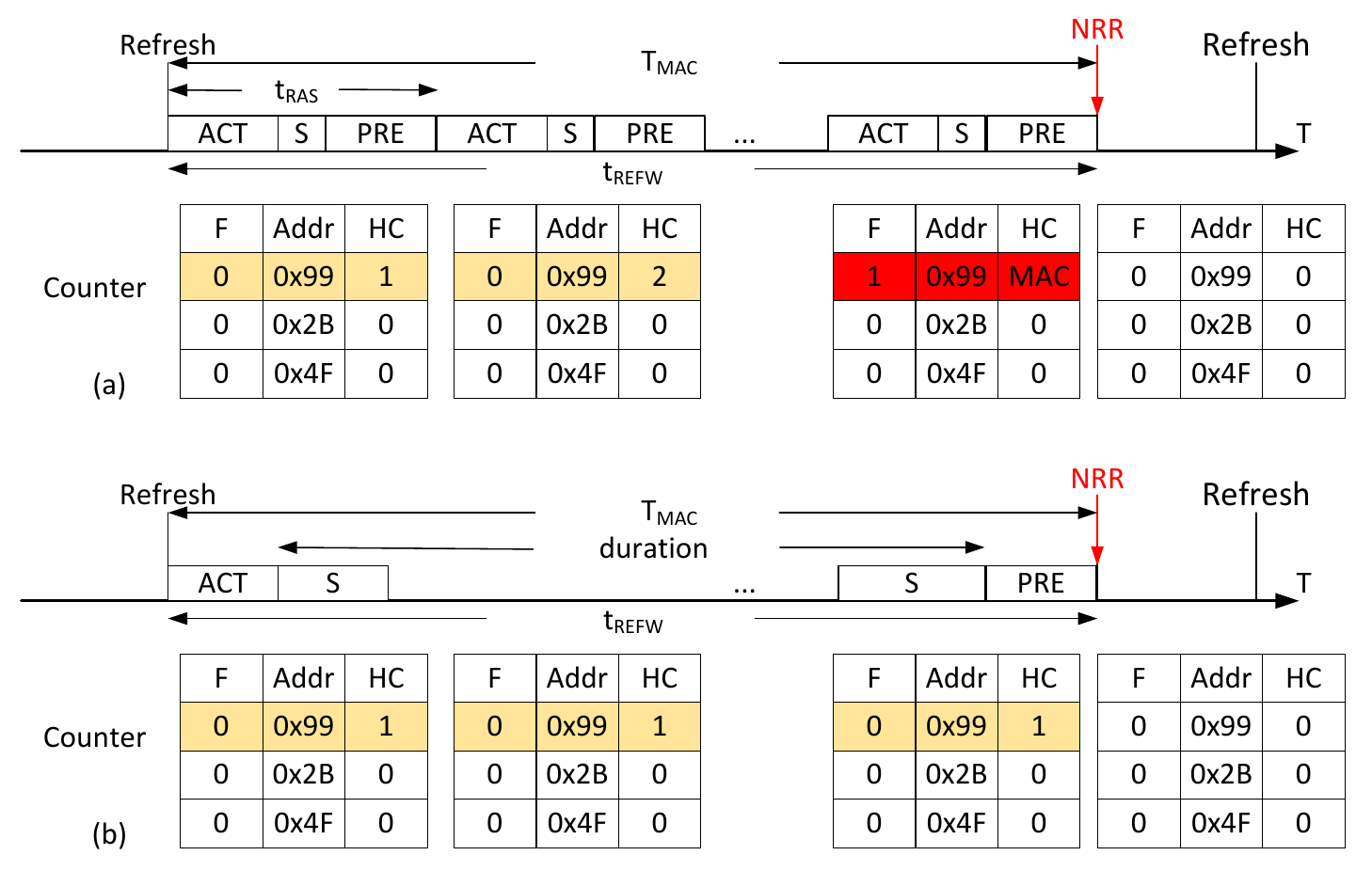}\vspace{-0.4em}
 \end{tabular} \vspace{-0.9em}
\caption{Timing of (a) RowHammer \& (b) RowPress Attack.}
\label{timing}
\vspace{-1.8em}
\end{center}
\end{figure}

Fig. \ref{timing}(b) shows the timing sequence of a Rowpress attack on row 0x99 as an example. This attack is implemented in such a way that the memory controller does not detect any anomaly because there is always only one activation command involved. 
Algorithm \ref{alg-RP} shows us step-by-step on how Rowpress works. It’s similar to Algorithm \ref{alg-RH}, and the main difference lies in line 2, where we take the time $T$ instead of the count $N$ as Input. Both algorithms use the same data pattern, but in Algorithm \ref{alg-RP}, we use one `ACT' and then wait until we can proceed, as shown in lines 6 to 9. After that, both algorithms do the same things to change and check the data.

\begin{algorithm}[b]
        \caption{\small RowPress Fault Injection}
          \scalebox{0.65}{
    \begin{minipage}{3\linewidth}
        \begin{algorithmic}[1]
            \State $\textbf{Procedure: \textit{CounterBypass}}$
                \State $\textbf{Input} \hspace{4pt} T$ 
                \State $Allocate \hspace{4pt} row\_address, bank\_address, column\_address$ 
                \State $Load \hspace{4pt} Data\_pattern \hspace{4pt} $\&$ \hspace{4pt} Data\_pattern\_inv $ 
                \State $Pattern\_row[row] \gets Data\_pattern\hspace{16pt}//We\hspace{4pt}assign\hspace{4pt}0xFFFFFFFF\hspace{4pt}to\hspace{4pt}Pattern\hspace{4pt}rows$
                \State $Victim\_row[row] \gets Data\_pattern\_inv\hspace{16pt}//We\hspace{4pt}assign\hspace{4pt}0x00000000\hspace{4pt}to\hspace{4pt}victim\hspace{4pt}rows$ 
                \State $\hspace{16pt}ACT \hspace{4pt} Pattern\_row[row];\hspace{16pt}//Activate\hspace{4pt}Row\hspace{4pt}X\hspace{4pt}once$ 
                \State $\hspace{16pt}Sleep(T)$
                \State $\hspace{16pt}PRE \hspace{4pt} Pattern\_row[row];$ 
        \State $\textbf{For}\hspace{4pt}(row\hspace{4pt} in\hspace{4pt}Row\_address)\hspace{4pt}\textbf{do}$
                    \State $\hspace{16pt}READ \hspace{4pt} Rows[row];$ 
                    \State $\hspace{16pt}PRE \hspace{4pt} Rows[row];$ 
                    \State $\hspace{16pt}row \gets row+1;$ 
        \State $Receive\_Data(Platform);\hspace{16pt}//Write\hspace{4pt}data\hspace{4pt}back\hspace{4pt}to\hspace{4pt}hostPC$
        \State $Detect\_BitFlips(Victim\_Row)$

    \State \textbf{end} $\textbf{Procedure}$
        \end{algorithmic} 
         \end{minipage} \vspace{-5em}}
         \label{alg-RP}
    \end{algorithm}

\vspace{-0.5em}

\section{DRAM-Profile-aware Attack} 
\begin{figure}[t]
\begin{center}
\vspace{-1em}
\begin{tabular}{c}
\includegraphics [width=0.75\linewidth]{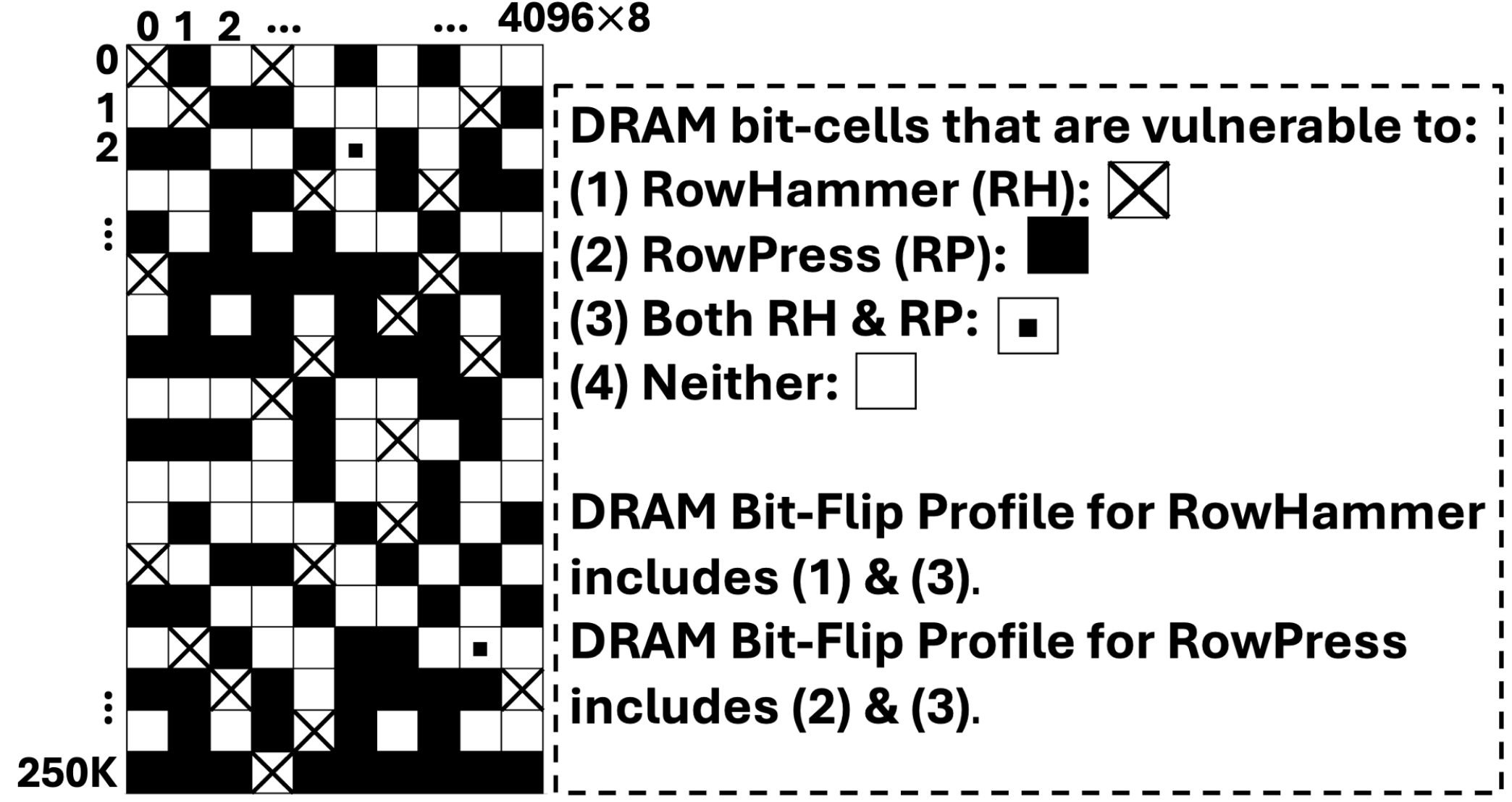}\vspace{-0.7em}
\end{tabular}
\caption{A schematic diagram of DRAM bit-cells illustrating the huge difference of vulnerable bit-cells under RowHammer (marked with cross) and RowPress (marked in solid black) attacks in terms of number and location. }
\label{DRAM-Profile}
\vspace{-2em}
\end{center}
\end{figure}

\vspace{-0.2em}
The first step for an attacker is to conduct DRAM profiling of the entire chip, leveraging the fault injection model of sections V.A and V.B for RowHammer and RowPress, respectively. After completing the profiling of the DRAM bit-cells under RowHammer and RowPress attacks, a sample DRAM chip will depict a schematic diagram of vulnerable bit-cell similar to Fig. \ref{DRAM-Profile}. Here, the cross cells are vulnerable to only RowHammer (RH), Black cells are vulnerable to only RowPress (RP), and Dot cells denote vulnerability to both RH and RP. Next, considering a DRAM chip with N cells to store the deep learning model, we can denote a subset of these cells as $C_{rh}$ and $C_{rp}$, which indicates the list of cell locations in the DRAM vulnerable to RowHammer and RowPress, respectively. After completing the profiling stage, an attacker will have a set of bit locations $C_{rh}$ or $C_{rp}$, where the bit-flip attack is feasible. These locations can identified by a unique page frame number and an offset. The next challenge would be leveraging these bit profiles to formulate an attack objective and search for appropriate vulnerable bit indexes to achieve the desired attack goal for deep learning models.




\subsection{Attack Objective}

Our proposed attack algorithm aims to deplete the intelligence of the DNN models. Taking classification as an example task, the attack objective would be to deteriorate the model accuracy close to a random guess level. For a quantized DNN with a bit precision of $n_\textup{q}$-the bit that can be represented in 2's complement binary Tensor as $\{B_l\}_{l=1}^L$, where $l \in \{1, 2,..., L\}$ is the layer index, the attack objective can be formulated as follows: \vspace{-0.7em}
\begin{equation}
\label{eqt:BFA}
\begin{gathered}
\max_{B}  ~\mathcal{L}\Big (f \big( x ; \{\hat{B}_l\}_{l=1}^{L} \big), {y} \Big) \\
\centering ~\small{\sum_{l=1}^{L}} ~\mathcal{D}(\hat{B}_l, B_l) \in \{0,1,...,n_b\},\vspace{-0.5em}
\end{gathered}
\end{equation} 
where $x$ and $y$ denote the selected input mini-batch and ground-truth labels; 
$f(x;\{\hat{B}_l\}_{l=1}^L)$ compute the outputs of DNN parameterized by $\{\hat{B}_l\}_{l=1}^L$; $\hat{B}$ denotes the model weight bits after attack; $\mathcal{L}(\cdot, \cdot)$ computes the cross-entropy loss.
The goal is to maximize the loss function $\mathcal{L}$ with the least number of bit-flips (i.e., minimizing $\mathcal{D}$); namely, minimizing attack overhead.
\vspace{-0.5em}
\subsection{Attack Algorithm}
An efficient vulnerable bit-search algorithm is needed to attain the above attack objective with minimal bit-flips. We adopt the popular adversarial weight attack algorithm, the BFA algorithm \cite{rakin2019bit}, for flipping targeted bits in DNN. BFA is a two-stage search algorithm that utilizes gradient ranking to search for vulnerable weight-bit indices. However, instead of naively applying the BFA algorithm, we incorporate our DRAM bit-flip profiles $C_{rh}$ or $C_{rp}$ appropriately in the search algorithm as shown in Algorithm \ref{alg-DNN}. We introduce a novel search strategy that considers the mapping of neural network weight-bits to bit-cell locations in a hardware DRAM profile.
For a quantized DNN represented as $\{B_l\}_{l=1}^L$, where $l \in \{1, 2,...,L\}$ is the layer index; initially, we select a subset of weight-bits, denoted as $\{B_{cl}\}_{l=1}^L \in \{B_l\}_{l=1}^L$, where $B_{cl} \in B_{l} ~\forall l$. $\{B_{cl}\}_{l=1}^L$ only contains weight-bits that can be mapped to a vulnerable bit-cell location in the DRAM bit-flip profiles: $C_{rh}$ or $C_{rp}$. In summary, before starting the vulnerable bit-search algorithm,  we confine our search within the weight-bits that are mapped to a vulnerable bit-cell location based on the hardware profiles. Note that, as an attacker, we do not alter or control the mapping between model weights and dram cells; rather, we only exploit the vulnerability in the existing mapping. As a result, the attack objective of eqn.~\ref{eqt:BFA} can be re-formulated as follows: \vspace{-1.5em}



\begin{equation}
\label{eqt:BFA_ref}
\begin{gathered}
\max_{B}  ~\mathcal{L}\Big (f \big( x ; \{\hat{B}_{cl}\}_{l=1}^{L} \big), {y} \Big) 
\end{gathered}
\end{equation}

Next, our proposed attack performs the conventional BFA search algorithm \cite{rakin2019bit} in an iterative manner. During each iteration, firstly, the intra-layer search identifies the bits with the highest gradients ( $|\nabla_{B_{cl}} \mathcal{L}|$) as vulnerable bit candidates. 
Then, the inter-layer search compares the bit candidates selected by the intra-layer search by looking at the loss values, i.e., the layer that induces the maximum loss will be elected, and the bits selected in that layer during the intra-layer search will be flipped. The attack continues to the next iteration until the objective is satisfied, as shown in Algorithm \ref{alg-DNN}.\vspace{-0.7em}

\begin{algorithm}[h]
    \caption{\small DRAM-profile-aware Attack}
    \scalebox{0.65}{
    \begin{minipage}{3\linewidth}
        \begin{algorithmic}[1]
            \State $\textbf{Procedure: \textit{DRAM-profile-aware Attack}}$
                \State Select a set of feasible weight-bits ($\{B_{cl}\}_{l=1}^L$) according to the DRAM bit-flip profiles: \\ $C_{rh}$ or $C_{rp}$.
                \State $\textbf{While}\hspace{4pt}Attack\hspace{4pt}Objective\hspace{4pt}is\hspace{4pt}not\hspace{4pt}Satisfied\hspace{4pt}\textbf{do}$
                    \State \hspace{16pt}$\textbf{For}\hspace{4pt}(l\hspace{4pt} <=\hspace{4pt}$L$)\hspace{4pt}\textbf{do}$
                        \State $\hspace{32pt}Find\hspace{4pt}vulnerable\hspace{4pt}weight\hspace{4pt}bit\hspace{4pt}with\hspace{4pt}highest\hspace{4pt}gradient\hspace{4pt} ( |\nabla_{B_{cl}} \mathcal{L}|)$
                        \State $\hspace{32pt}Perform\hspace{4pt}bitflip\hspace{4pt}to\hspace{4pt} record\hspace{4pt}loss\hspace{4pt}(\mathcal{L}\Big (f \big( x ; \{\hat{B}_{cl}\}_{l=1}^{L} \big), {y} \Big) \hspace{4pt}and\hspace{4pt}bit\hspace{4pt}index $
                        \State $\hspace{32pt}Restore\hspace{4pt}the\hspace{4pt}bit\hspace{4pt}back$ 
                    \State \hspace{16pt}$Enter\hspace{4pt}the\hspace{4pt} layer\hspace{4pt}{l_m}\hspace{4pt}with\hspace{4pt}maximum\hspace{4pt}loss\hspace{4pt}\mathcal{L} $
                    \State \hspace{16pt}$Perform\hspace{4pt}bitflip\hspace{4pt} at\hspace{4pt}layer\hspace{4pt}l_m\hspace{4pt}on\hspace{4pt}the\hspace{4pt}recorded\hspace{4pt}index$
    \State \textbf{end} $\textbf{Procedure}$
    \end{algorithmic} 
    \end{minipage} 
    }
    \label{alg-DNN}
\end{algorithm}

\vspace{-0.5em}
\section{Experiment Results}
\subsection{Hardware Setup}
\begin{figure}[t]
\begin{center}
\begin{tabular}{c}
\includegraphics [width=0.85\linewidth]{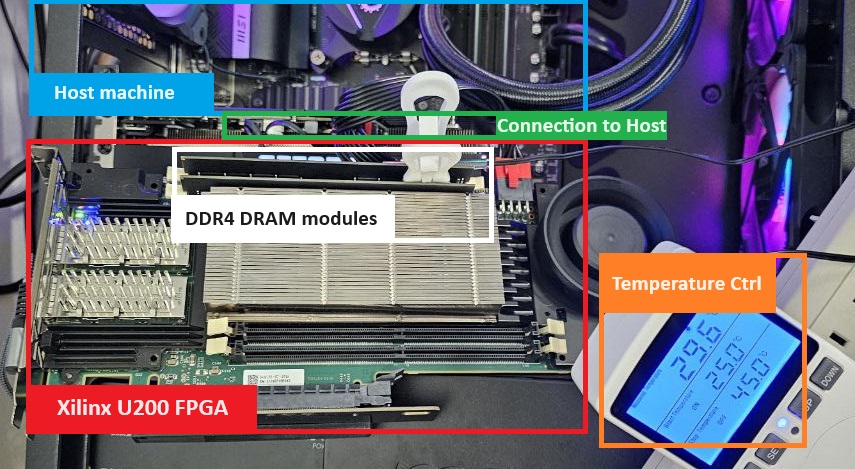}
 \end{tabular} 
\caption{Our testing infrastructure for DDR4 modules.}
\label{frame}
\vspace{-1.8em}
\end{center}
\end{figure}

\noindent\textbf{DRAM Testing Infrastructure.}
We extensively modify the DRAM-Bender \cite{olgun2023dram} to have a versatile FPGA-based DRAM attack exploration framework for DDR4 with an in-DRAM compiler API installed on our host machine. Our testing infrastructure (as Fig. \ref{frame} illustrates) consists of the Alveo U200 Data Center Accelerator Card \cite{Alevo}, which serves as the FPGA that accepts DDR4 modules and runs the test programs based on Algorithms \ref{alg-RH} \& \ref{alg-RP} via sending DDR4 command traces generated by the host machine. The key idea is to take control of memory modules for DDR4 interfaces with straightforward high-level programming to test, characterize, and run the generated programs on the host machine. The driver is designed to send instructions across the PCIe bus to the FPGA to be stored on the board. The temperature is kept below 30$^{\circ}$C with INKBIRDPLUS 1800W temperature controller.

\noindent\textbf{Minimizing Interference.} To ensure that we directly observe RowHammer and RowPress's circuit-level bit-flips, DRAM refresh \cite{JEDEC} and rank-level ECC are disabled. However, proprietary RowHammer protection techniques (e.g., Target Row Refresh \cite{frigo2020trrespass,hassan2021uncovering}) exist. 

\noindent\textbf{Tested Commodity DDR4 DRAM Chip.}
We select one representative Samsung-manufactured DRAM chip with 16GB density (Frequency: 2400MHz, Die revision: B, Org.: x8, Date: 2053) to profile its cell vulnerabilities.

\begin{table*}[t]
\centering
\caption{Results of RowHammer \& RowPress attacks on different applications, datasets and DNN architectures. We report the number of bit-flips required to degrade the DNN performance back to a random guess level.} 
\scalebox{0.74}{
\begin{tabular}{ccccccccc}
\hline
\multicolumn{1}{c}{\textbf{Dataset}} & \textbf{Architecture} & \textbf{\#Parameters} & \textbf{\thead{Acc. before \\ Attack (\%)}}  & \textbf{\thead{Random Guess \\ Acc. (\%)}} & \textbf{\thead{Acc. After \\ RowHammer Attack (\%)}} & \textbf{\thead{\#Bit-flips \\ (RowHammer Attack)}} & \textbf{\thead{Acc. After \\ RowPress Attack (\%)}} & \textbf{\thead{\#Bit-flips \\ (RowPress Attack)}}\\ \hline
\multirow{3}{*}{CIFAR-10} & ResNet-20 & 0.27M & 92.42 & \multirow{3}{*}{10.00} & 10.39 & 36 & 9.14 & 8 \\
& ResNet-32 & 0.47M & 93.44 & & 10.41 & 60 & 10.28 & 11 \\
& ResNet-44 & 0.66M & 93.90 & & 10.4 & 53 & 10.47 & 14 \\
\hline
\multirow{7}{*}{ImageNet} & ResNet-34 & 21.8M & 73.12 & \multirow{7}{*}{0.10} & 0.14 & 35 & 0.13 & 11 \\
& ResNet-50 & 25.6M & 75.84 & & 0.11 & 26 & 0.13 & 10 \\
& ResNet-101 & 44.6M & 77.20 & & 0.14 & 30 & 0.14 & 11 \\
& DeiT-T & 5.7M & 71.95 & & 0.15 & 143 & 0.12 & 45 \\
& DeiT-S & 22M & 79.63 & & 0.15 & 56 & 0.07 & 24 \\
& DeiT-B & 86.6M & 81.7 & & 0.14 & 47 & 0.13 & 13 \\
& VMamba-T & 23M & 81.82 & & 0.12 & 79 & 0.12 & 24 \\
\hline
\thead{Google Speech Command} & M11 & 1.8M & 93.2 & 2.86 & 2.84 & 68 & 2.44 & 19\\
\hline
\end{tabular}}
\vspace{-1em}
\label{results} 
\end{table*}

\noindent\textbf{Fair Evaluation Settings for RowHammer \& RowPress.}
When comparing the number of bit-flips, we use hammer counts (HCs) for RowHammer and the number of cycles elapsing within an activation duration for RowPress, both can be converted into time for a fair comparison. For instance, a 2400MHz DRAM chip with 100 million cycles takes approximately $T=\frac{100M}{2400M}=41.67ms$. The equivalent HCs in RowHammer can be calculated as $HC=\frac{T}{t_{REF}}\times1.36M=\sim885.5K$, with $t_{REF}$ typically being $64ms$ \cite{liu2013experimental}. 

\vspace{-0.5em}
\subsection{Deep Learning Framework}
\noindent\textbf{DNN Architectures \& Datasets.} 
We evaluate both RowHammer and RowPress attacks on vision and speech applications. For image classification tasks, we have ResNet-\{20,32,44\} \cite{he2015delving} trained on CIFAR-10 \cite{krizhevsky2010cifar}, as well as large-scale DNNs trained on ImageNet \cite{deng2009imagenet}, including ResNet-\{34,50,101\} \cite{he2015delving}, DeiT-\{T,S,B\} \cite{touvron2021training}, and VMamba-T \cite{liu2024vmamba}. For speech recognition, we have the M11 model \cite{dai2017very} trained on Google's speech command dataset \cite{warden2018speech}. We perform an 8-bit post-training quantization for all the aforementioned models following \cite{rakin2019bit, rakin2021deep}. In addition, we run our attack experiments three times to report their average, reducing the impact of random attack initialization (e.g., random test batch selection, and mapping of weights to vulnerable bit-cells). 

\noindent\textbf{Evaluation Metric.} We report the number of required bit-flips to degrade the model accuracy close to a random guess level (i.e., $\frac{1}{\#\text{output classes}} \times 100\%$). A lower number of bit-flips indicates higher attack efficiency. 

\begin{figure}[t]
\begin{center}
\begin{tabular}{c}
\includegraphics [width=0.9\linewidth]{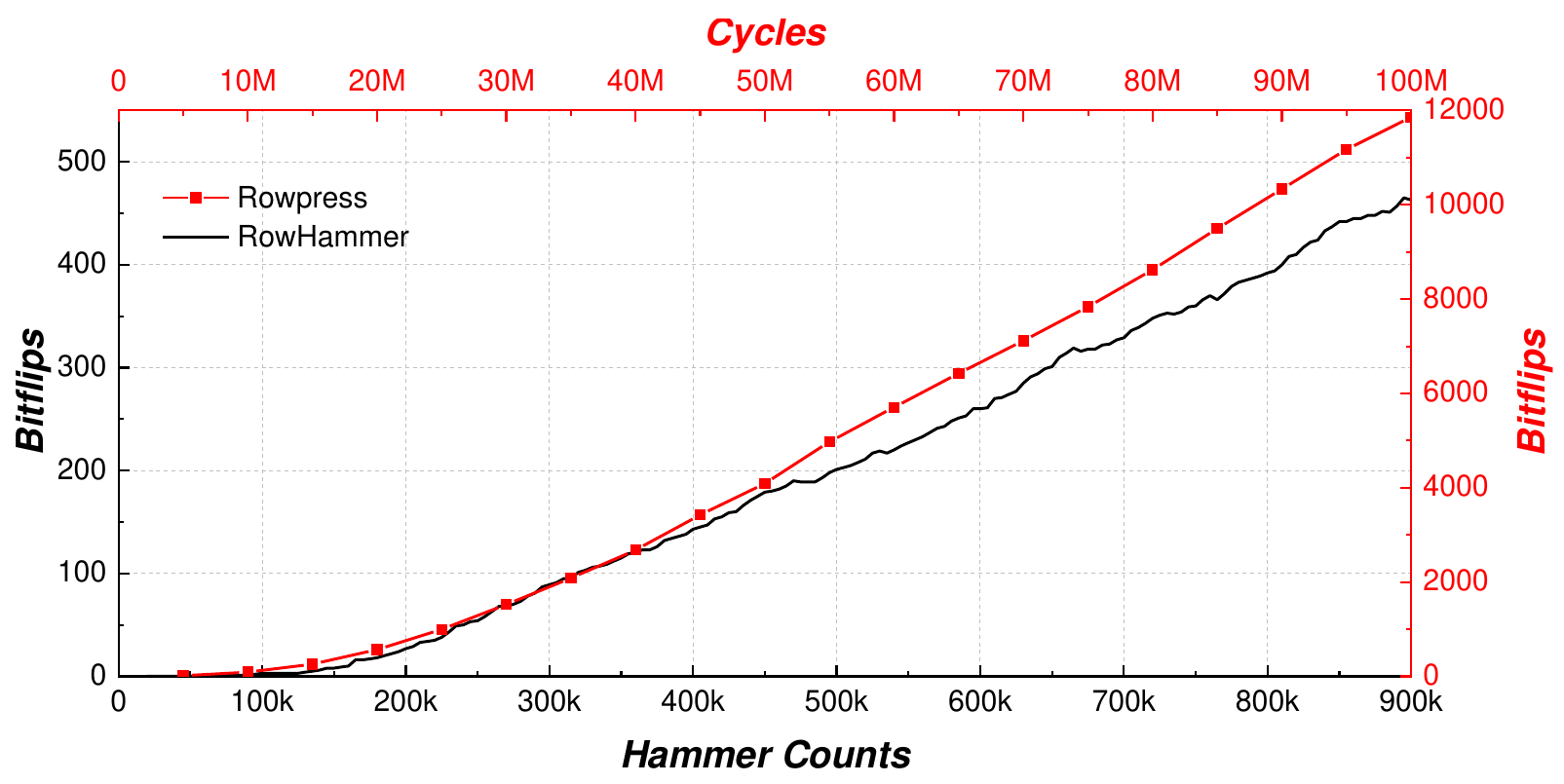}\
 \end{tabular} \vspace{-1em}
\caption{Experimental results on double-sided RowHammer attack vs. RowPress.}
\label{RHRP}
\vspace{-2em}
\end{center}
\end{figure}

\subsection{Results: RowHammer vs. RowPress}
\subsubsection{DRAM Vulnerability Analysis}
As shown in Fig. \ref{RHRP}, the outcomes of the two distinct attack models are depicted in red and black curves. The red curve, associated with the cycle counts and bit-flips axes on the top and right quantifies the bit-flips induced by RowPress as a function of cycle counts. The black curve, associated with the hammer counts and bit-flips axes on the bottom and left, demonstrates the correlation of bit-flips with the increasing number of hammer counts ascribed to the RowHammer attack. This figure indicates that both attack vectors result in an increase in bit-flips over time. Notably, the red curve predominates over the black one for the majority of the observational period, with RowPress producing 20$\times$ more bit-flips than RowHammer does.

\vspace{0.5em}
\hspace{-0.5em}\fcolorbox{black}{white}{\begin{minipage}{24em}
\textbf{Takeaway 1.} \emph{Given a similar attack budget (i.e., resources such as time), RowPress produces 20$\times$ more bit-flips than RowHammer.} 
\end{minipage}}\vspace{0.5em}

Echoing \cite{luo2023rowpress}, this crucial observation suggests that RowPress presents a more pernicious form of attack, capable of inflicting more extensive memory disruption.

\subsubsection{DRAM-profile-aware Attack Evaluation}

\noindent\textbf{\\ Evaluation of RowPress on Compromising DNNs.}
Table \ref{results} summarizes the performance evaluation of our DRAM-profile-aware attack, exhibiting that DNNs of various forms are extremely vulnerable to RowPress. We evaluated ten vision models consisting of three distinct architectural topologies: CNN, vision transformer, and VMamba. CNNs, represented by ResNets, are noticeably more vulnerable than vision transformers and Vmamba, as their intelligence can be depleted with much fewer attack iterations. RowPress also presents effective attack potency on the audio classification model, M11. As shown in the last column of Table \ref{results}, the RowPress attack requires no more than 45 bit-flips and averages 18 bit-flips to deplete the intelligence of all tested DNN models.  

\begin{figure}[h]
\begin{center}
\begin{tabular}{c}
\includegraphics [width=0.9\linewidth]{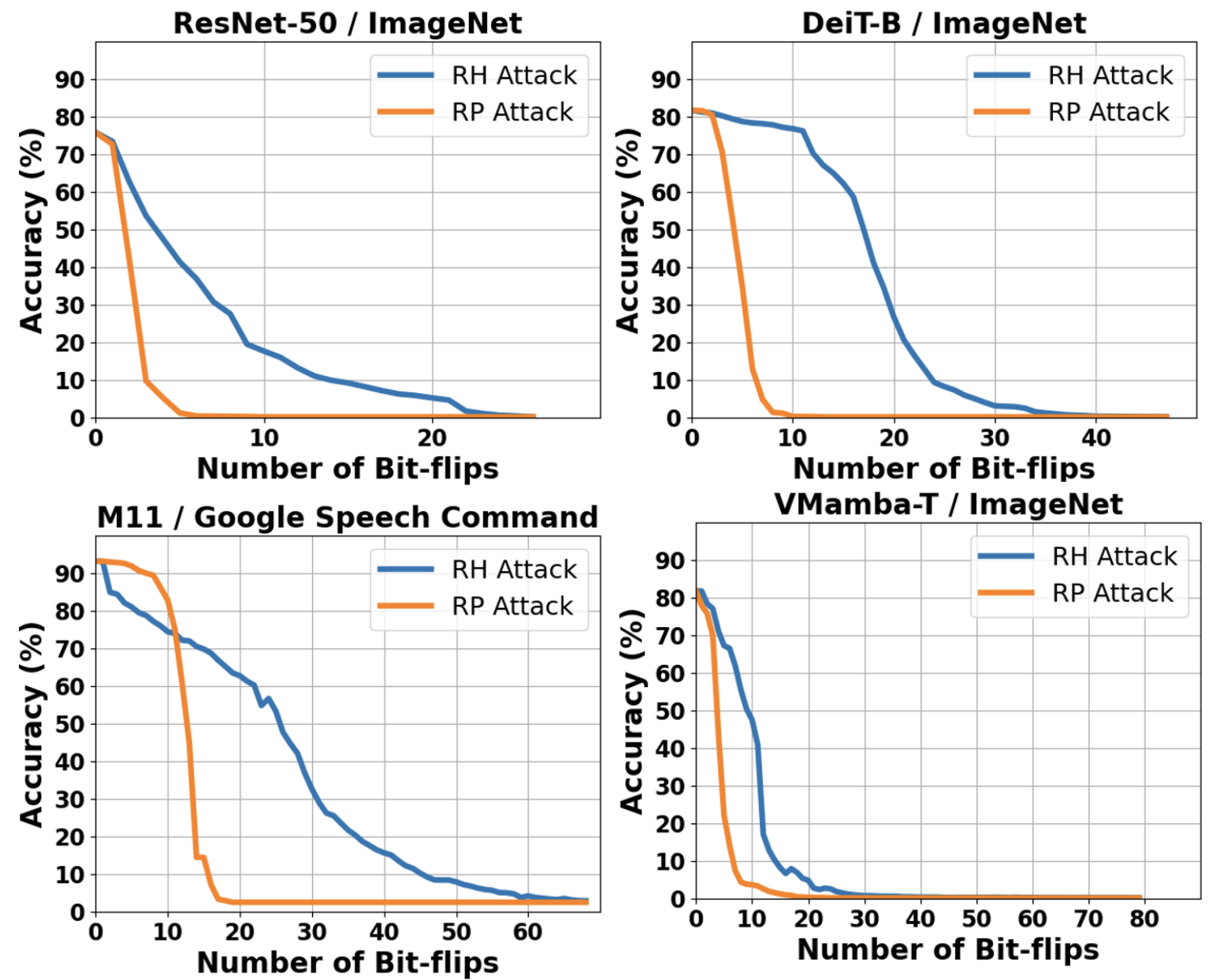} 
\end{tabular}
\caption{Accuracy evolution curves versus the number of bit-flips under RowHammer (RH) and RowPress (RP) attacks.}
\label{acc-curves} 
\end{center}
\end{figure}

\vspace{0.5em}\hspace{-0.5em}\fcolorbox{black}{white}{\begin{minipage}{24em}
\textbf{Takeaway 2.} \emph{RowPress is highly effective in depleting the intelligence of DNN models of various sizes, topology, training data modality and resolution.}
\end{minipage}}\vspace{0.5em}

\noindent\textbf{RowHammer vs. RowPress Analysis.}
Fig. \ref{acc-curves} presents some representative DNN accuracy degradation curves under our DRAM-profile-aware attack. We observe that the slopes of orange curves (corresponding to the RowPress bit-flip profile) are noticeably steeper than those of blue curves (corresponding to the RowHammer bit-flip profile). We can reasonably attribute this phenomenon to the fact that the DRAM bit-flip profile of RowPress contains more vulnerable bits than that of RowHammer, as Fig. \ref{DRAM-Profile} illustrates. \textit{Quantitatively}, the RowPress bit-flip profile contains \textit{more} vulnerable bits; \textit{qualitatively}, bits contained in the RowPress profile are more vulnerable and produces more negative and disastrous effect on the model once they are flipped. This twofold property contributes to RowPress's superior attack efficiency. Both fault injection methods exhibit roughly similar attack efficacy trends across different model sizes and structures. The performance gap is largest for DeiT-B, whereas the gap for VMamba-T is relatively small. Prior works~\cite{mahmood2021robustness,zhang2023transferable} have shown that vision transformers are more resilient against perturbations, which supports our observations.

\vspace{0.5em}
\hspace{-0.5em}\fcolorbox{black}{white}{\begin{minipage}{24em}
\textbf{Takeaway 3.} \emph{RowPress exhibits noticeably higher attack efficiency compared to RowHammer, averaging 3.6$\times$ fewer bit-flips to attain the same attack objective. }
\end{minipage}}


\section{Conclusion}
In this paper, we present a pioneering investigation into the potential impact of RowPress on executing targeted bit-flip attacks against deep neural networks. Our analysis reveals that RowPress enables an attacker to degrade the performance of DNNs with unprecedented efficiency, surpassing the efficacy of the RowHammer attack. We discover that, within the same operational duration, RowPress can induce up to twenty times more bit-flips in a DRAM compared to RowHammer. When a DNN is deployed on the DRAM, to deplete the DNN's intelligence to a random-guess level, our DRAM-profile-aware attack requires up to four times fewer bit-flips with the RowPress bit-flip profile than with the RowHammer bit-flip profile. We have implemented RowPress on physical DRAM chips and rigorously evaluated its impact on the performance of various DNN models. The results confirm that bit-flips executed via RowPress can significantly degrade the performance of all tested models. Therefore, based on our findings, it is crucial for the AI community to address the security threat posed by RowPress by investigating into appropriate protective measures.

\bibliographystyle{IEEEtran}
\bibliography{main.bib}

\end{document}